\documentclass[prd,aps,showpacs,groupedaddress,eqsecnum,notitlepage,nofootinbib,onecolumn,superscriptaddress,twocolumn]{revtex4-2}
\usepackage{bbm}
\usepackage{graphicx}
\usepackage{tikz,siunitx,mwe}
\usepackage{tensor}
\usepackage{epstopdf}
\usepackage{amsmath}
\usepackage{amsfonts}
\usepackage{slashed}
\usepackage{amssymb}
\usepackage{enumerate, color}
\usepackage{subfigure}
\usepackage{lipsum}
\usepackage{color}
\usepackage{graphicx,bm}
\usepackage[utf8]{inputenc}
\usepackage{eurosym}\usepackage{scalerel}
\usepackage{float}
\usepackage{accents}
\usepackage[colorlinks=true,pdfstartview=FitV,linkcolor=blue,citecolor=blue,urlcolor=blue,breaklinks=true]{hyperref}
\usepackage{orcidlink}
\newcommand{\be}{\begin{equation}}
\newcommand{\ee}{\end{equation}}
\newcommand{\ben}{\begin{eqnarray}}
\newcommand{\een}{\end{eqnarray}}
\newcommand{\bes}{\begin{subequations}}
\newcommand{\ees}{\end{subequations}}
\usepackage{comment}
\def\bal#1\eal{\begin{align}#1\end{align}}

\def\bal#1\eal{\begin{align}#1\end{align}}
\begin{document}
\title{Zero modes and geometric phase for 2D Weyl fermions on Lifshitz backgrounds}
\author{G. Q. Garcia\,\orcidlink{0000-0003-3562-0317}}
\email{gqgarcia99@gmail.com}
\affiliation{Centro de Ciências, Tecnologia e Saúde, Universidade Estadual da Paraíba, 58233-000, Araruna, PB, Brazil}
\author{D. C. Moreira\,\orcidlink{0000-0002-8799-3206}}\email{moreira.dancesar@gmail.com}\affiliation{Centro de Ciências, Tecnologia e Saúde, Universidade Estadual da Paraíba, 58233-000, Araruna, PB, Brazil}
\author{E. Cavalcante\,\orcidlink{0000-0003-0953-0078}}\email{everton@servidor.uepb.edu.br}\affiliation{Centro de Ciências e Tecnologia, Universidade Estadual da Paraíba, 58429-500, Campina Grande, PB, Brazil}
\author{C. Furtado\,\orcidlink{0000-0002-3455-4285}}
\email{furtado@fisica.ufpb.br}
\affiliation{Departamento de F\'isica, Universidade Federal da Para\'iba, 58051-970, Jo\~ao Pessoa, PB, Brazil. }

\begin{abstract}

Here we investigate analytical properties of Weyl fermions in (2+1)-dimensional Lifshitz spacetimes. In particular, we are interested in obtaining geometric phases and verifying the existence of well-behaved fermionic zero modes. Using the Dirac phase method, we show how geometric phases naturally arise from the coupling between the fermionic fields and the Lifshitz geometry. We also present exact solutions of the zero modes by rewriting the Weyl equation as a system of supersymmetric equations.
\end{abstract}
\maketitle
\section{Introduction}
The behavior of matter fields in curved spacetimes has been a subject of study in Physics for a long time and has played a role as a bridge for several applications of gravitational models on condensed matter systems and {\it vice versa}. Within this context, the versatility of a class of materials known as Weyl semimetals has made it possible to study electronic properties of fermions in a large number of scenarios \cite{vafek2014dirac, hosur2013recent}. This class of semimetals exhibits a linear dispersion relation, where the conduction band touches the valence band at Weyl nodes with well-defined, distinct chirality. Particularly in the last decade, it has been experimentally verified that several materials exhibit properties of Weyl semimetals. The most famous of these is graphene, formed by a strictly two-dimensional hexagonal malleable structure of carbon atoms discovered by Geim and Novoselov~\cite{geim2007rise},  which under specific conditions, exhibits behavior reminiscent of a Weyl semimetal despite being formally classified as a Dirac material~\cite{katsnelson2012graphene}. Its malleability and the presence of topological defects within its lattice structure allows us to study curvature effects such as disclinations and dislocations, which may emerge separately or simultaneously. For instance, graphene has been described in different curved backgrounds presenting spherical~\cite{gonzalez1993electronic, garcia2017geometric}, cylindrical~\cite{ando2005theory, ando2009electronic}, helical~\cite{watanabe2015electronic, atanasov2015helicoidal} and some others~\cite{bueno2012landau, garcia2025landau, garcia2025rotation, silva2024strain, CavalcanteFurtado2024, bakke2012kaluza} symmetries. Another Weyl semimetal can be found at the superfluid phase of $^{3}He-A$, which can simulate gravitational effects as cosmic string, axial anomaly, vacuum polarization and others ~\cite{volovik1998simulation, carvalho2025scaterring}. Consequently, this class of materials provides excellent laboratories for exploring different effects of geometry in condensed matter models and, furthermore, by applying these ideas to study different elements such as pyrochlore iridates \cite{PhysRevB.83.205101, matsuhira2007metal, taira2001magnetic}, $Bi_2Se_3$ compound~\cite{PhysRevLett.107.127205, zhang2009topological, xia2009observation} and  $HgTe$~\cite{bernevig2006quantum}, new technological applications may emerge. 

A way to enforce this possibility of exploring different effects of background geometries on matter fields in the search for realistic applications can be made through gauge/gravity duality, where isometries of bulks in gravitational models map symmetries of fields located at the bulk boundaries \cite{ammon2015gauge,hartnoll2009lectures}. Originally addressed to the analysis of relativistic models, in a subject that is now known as the AdS/CFT Correspondence \cite{maldacena1999large,aharony2000large} (connecting in the same setup bulk properties of asymptotically Anti-de Sitter (AdS) spacetimes with Conformal Field Theories (CFTs) living on its boundary), the gauge/gravity duality has been extended on several areas and is understood as a powerful tool for analyzing strongly correlated and critical systems (For instance, see \cite{hartnoll2009lectures} and related references). In particular, in order to study nonrelativistic strongly coupled systems in condensed matter, {\it Lifshitz spacetimes} were proposed as gravity duals in applications involving quantum critical points \cite{kachru2008gravity}, but it was soon understood that the fermion behavior on these backgrounds could also be used to study Fermi surfaces \cite{faulkner2011semi} and strange metals \cite{hartnoll2010towards}. In addition, several analytical properties of fermion field solutions on Lifshitz backgrounds, including Green's functions, were presented and explored in \cite{alishahiha2012fermions,gursoy2012holography}. What allows these spaces to provide gravity descriptions of nonrelativistic models is the fact that the temporal coordinate has a scaling property different from the one presented by spatial coordinates, which induces an scaling anisotropy in the background metric generated by the field equations. Such anisotropy is measured by a {\it dynamical exponent} which can be adjusted to map the exact scaling property to be satisfied by quantum critical points of the field model living on the spacetime boundary. 

One can use semi-holographic \cite{gursoy2013holographic} and holographic \cite{landsteiner2016quantum,landsteiner2020holographic} methods to explore properties of Weyl semimetals from gravity descriptions in Lifshitz spacetimes. Indeed, in these setups, asymptotically Lifshitz black branes in $4+1$ dimensions are used to provide gravity descriptions of strongly coupled fermions on its respective $3+1$ dimensional boundaries. The dual field theory which emerges is nonrelativistic due to the inherent anisotropic scaling symmetry and, by addressing the problem to study Weyl fermions, we find a scenario where the boundary of the system behaves similarly to Weyl semimetals, since in these materials the Fermi velocity is small compared to the speed of light and in the vicinity of the Weyl nodes the emerging quasiparticles are effective massless fermions with well defined chirality. From this perspective, phenomena such as anomalous transport, chiral magnetic effects, and topological phase transitions have been analyzed from a holographic viewpoint \cite{ammon2015gauge, Ahn2024_Praper, palumbo2016holographic}.  Additionally, the recent experimental realization of two-dimensional Weyl semimetals \cite{lu2024realization} highlights the need to expand our understanding on the behavior of massless fermions on curved two-dimensional surfaces, as well as their links with different physical systems. These surfaces can be understood as the boundary of an extremal 3+1 black brane model equipped with scaling anisotropy and studying the behavior of fields on these surfaces can give us more hints about the physics of strongly coupled non-relativistic quantum systems.

 The discussion above indicates the existence of a close relation between the physics of Weyl fermions and the properties of Lifshitz spacetimes in the description of certain condensed matter phenomena. Therefore, thinking along these lines, one way to better understand this relationship is through the study of massless fermions on background geometries which exhibit anisotropic scaling isometries. Such analyses can provide new hints about the physics of these new materials, helping to better understand them and opening new avenues of investigation. Here we are interested in studying some properties of Weyl fermions on $2+1$-dimensional Lifshitz backgrounds. In particular, we show that on these setups the Weyl spinor acquires a geometric phase and presents a well-behaved zero mode, derived from an underlying supersymmetric quantum mechanics. Holonomy calculus studies the transport of a given quantity to the spin connection. This quantity depends on the path geometry and, for closed and adiabatic paths, it is usually called Berry geometric phase. But in a spacetime, we have several possibilities of closed paths, which involve all coordinates - time included. When one consider a given path involving a displacement in the time coordinate, a non-trivial holonomy associated with temporal translations is captured. This phase arises from the generalization on holonomy calculus, i.e., it is a generalization of the Berry phase \cite{bezerra1992,gomes2007loop,de2003loop,de2025adiabaticity,cai1991neutrino}. The physical interpretation of this temporal holonomy is associated with a shift in energy due to time boosts associated to chemical potential and the emerging squeezed quantum states. 

This work is organized as follows. In Sec. II, we derive the Weyl equation in a 2+1-dimensional Lifshitz background and present the corresponding spin connections. In Sec. III we calculate the geometric through the Dirac phase method and the holonomy matrix is presented as a series expansion. In Sec. IV, we identify the fermionic zero modes and explore their underlying supersymmetric structure. Finally, in Sec. V, we summarize our conclusions and discuss potential implications.

\section{Weyl equation on Lifshitz spacetime}

In this paper we deal with Weyl fermions in 2 + 1 dimensions living on a Lifshitz spacetime with background geometry given by
\begin{eqnarray}\label{lifmetric}
ds^2 = -\left(\frac{\ell}{u}\right)^{2z} dt^2 + \left(\frac{\ell}{u}\right)^{2} \left(du^{2} + dx^{2}\right),
\end{eqnarray}
where $\ell$ is a length scale related with the spacetime curvature, $z$ is the {\it dynamical exponent} and $u$ denotes a radial coordinate. The line element \eqref{lifmetric} is invariant under the anisotropic scaling $\left(t,u,x\right)\rightarrow\left(\beta^z t,\beta u, \beta x \right)$ and has several applications in nonrelativistic gauge/gravity models, since it represents the gravity side of models dual to quantum systems with Lifshitz invariance and energy scale set by $u$ \cite{hartnoll2009lectures}. In particular, for $z=2$ the background metric \eqref{lifmetric} is dual to Lifshitz field theory and for $z=3$ it is possible to find consistent proposals of well-behaved (renormalizable) gravity in the UV regime of Hor\"ava-Lifshitz (HL) models \cite{taylor2008non,taylor2016lifshitz,zaanen2015holographic,hovrava2009quantum,sotiriou2011hovrava}. Furthermore, still for $z=2$, the same scaling property can be found when dealing with nonrelativistic fermions in flat spacetime, whose isometries correspond to the Schrödinger group in the gravity dual description \cite{akhavan2009fermions}. The relativistic scenario is recovered for $z\to 1$, where we retrieve the usual AdS spacetime. We can factorize the metric \eqref{lifmetric} as $g_{\mu\nu}=e^a_{~\mu}(x)e^b_{~\nu}(x)\eta_{ab}$, where $\left(x^0,x^1,x^2\right)=\left(t,u,x\right)$ and
\bes\ben
e^{a}_{\,\,\,\mu}\!\left(x\right)&=& \frac{l}{u}\!\left(
\begin{array}{cccc}
\left(\frac{\ell}{u}\right)^{\!z-1}\cosh{u} & \sinh{u} & 0\\ 
\left(\frac{\ell}{u}\right)^{\!z-1}\sinh{u} & \cosh{u} & 0\\
 0 & 0 & 1\\ 
\end{array}\right)\!,~~\\[6pt]
e^{\,\,\,\mu}_{a}\!\left(x\right)&=& \frac{u}{\ell}\!\left(
\begin{array}{cccc}
 \left(\frac{u}{\ell}\right)^{\!z-1}\cosh{u} & -\left(\frac{u}{\ell}\right)^{\!z-1}\sinh{u} & 0\\ 
-\sinh{u} & \cosh{u} & 0\\
 0 & 0 & 1\\ 
\end{array}\right)\!,~~~~~~~
\label{3.1.4}
\een\ees
 represents the tetrad and its inverse, respectively. One can also calculate the nonzero 1-form connections, listed in Table \eqref{tab1}, through the Maurer–Cartan relation $d\hat{\theta}^a + \omega^{a}_{\ b}\wedge\hat{\theta}^{b} = 0$, where the torsion 2-form is zero. 
\begin{table}[t!]
\centering
\caption[ ]{Spin connection components}
\begin{tabular}{ccc}
\hline
\toprule
\begin{minipage}[t]{.225\textwidth} \centering {\bf Components} \end{minipage}&
\begin{minipage}[t]{.225\textwidth} \centering {\bf Connections} \end{minipage}\\
\hline
\\[0.0005em]
$\omega^{\ 0}_{t\ 1} = \omega^{\ 1}_{t\ 0}$ & $-z\frac{\ell^{z-1}}{u^z}$ \\[10pt]
$\omega^{\ 0}_{u\ 1} = \omega^{\ 1}_{u\ 0}$ & $-1$  \\[10pt]
$\omega^{\ 0}_{x\ 2} = \omega^{\ 2}_{x\ 0}$ & $-\frac{\sinh{u}}{u}$  \\[10pt]
$\omega^{\ 1}_{x\ 2} = -\omega^{\ 2}_{x\ 1}$ & $-\frac{\cosh{u}}{u}$ 	 \\[10pt]
\hline
\toprule
\end{tabular}
\label{tab1}
\end{table}

The Weyl equation in (2+1)-dimensions is given by
\begin{equation}\label{weq}
 i\sigma^{\mu} \nabla_{\mu}\psi = 0,
\end{equation}
where the matrices \(\sigma^3=\gamma^0, i\sigma^1=\gamma^1, i\sigma^2=\gamma^2\) are set to respect the Clifford algebra \(\{\gamma^{\mu},\gamma^{\nu}\}=2\eta^{\mu \nu}\mathbb{I}\) and \(\nabla_{\mu}\) denotes the covariant derivative. In curved backgrounds, each $\sigma^{\mu}$-matrix can be expanded in the tetrad basis as \(\sigma^{\mu}=\sigma^{a} e_{a}^{\ \mu}(x)\) and for the metric \eqref{lifmetric} we have
\bes\ben
\sigma^{t}&=& \left(\frac{u}{\ell}\right)^z \left( \sigma^3 \cosh (u) -  i \sigma^1 \sinh (u) \right),\\
\sigma^{u}&=& - \frac{u}{\ell}  \left(\sigma^3 \sinh (u) - i \sigma^1 \cosh (u) \right),\\
\sigma^{x}&=&  \frac{u}{\ell} ( i \sigma^2),
\een\ees
with \(\{\sigma^{a}\}=\{\sigma^{t},\sigma^{u},\sigma^{x}\}\). In this way, the covariant derivative becomes  \cite{birrel1982quantum}
\begin{equation}
\nabla_{\mu}=\partial_{\mu}+\Gamma_{\mu}(x)=\partial_{\mu}+\frac{i}{4}\omega_{\mu ab}\Sigma^{ab}\mbox{,}
\end{equation}
with \( \Sigma^{ab}=\frac{i}{2}[\sigma^a, \sigma^b] \),
and a direct calculation reveals that the associated spin connections \(\Gamma_{\mu}(x)\) are 
\bes\ben
\Gamma_{t}(x)&=&\frac{z}{\ell}\frac{\ell^{z}}{u^z}\frac{\sigma^2}{2},\\
\Gamma_{u}(x)&=&\frac{\sigma^2}{2},\\
\Gamma_{x}(x)&=&-\frac{i}{2u} \big ( \sigma^1 \sinh{u} + i \sigma^3 \cosh{u} \big ).
\een\ees
Therefore, the Weyl equation (\ref{weq}) becomes
\begin{equation}\label{weyl equation}
i\!\left(\!\sigma^t \bigg ( \frac{\partial}{\partial t} \!+\! \frac{z}{\ell}\frac{\ell^{z}}{u^z}\frac{\sigma^2}{2} \bigg )\!+\!\sigma^u \bigg ( \frac{\partial}{\partial u} \!+\! \frac{1}{2u}\!+\!\frac{\sigma^2}{2} \bigg)\!+\!\sigma^x \frac{\partial~}{\partial x} \right)\!\psi\!= \!0.~~~~
\end{equation}
In order to find a solution to the Weyl equation in the expression above, one must perform a similarity transformation on the curved Dirac matrices \( \gamma^\mu\) \cite{villalba2001energy, bueno2012landau}.  This transformation is unitary, with \( A^{-1}(u) A(u) = \mathbbm{1} \), and given by
\begin{equation}
    A(u) = exp \left(i\frac{u}{2} \gamma^{2}\right),
\end{equation}
leading us to the relations
\begin{subequations}
\begin{eqnarray}
    A^{-1}(u) \sigma^{t} A(u) = \left(\frac{u}{\ell}\right)^{z} \sigma^{3},\\
    A^{-1}(u) \sigma^{u} A(u) = \frac{u}{\ell} \left(i\sigma^{1}\right),\\
    A^{-1}(u) \sigma^{x} A(u) = \frac{u}{\ell} \left(i\sigma^{2}\right).
\end{eqnarray}
\end{subequations}
With all these ingredients, the equation \eqref{weyl equation} assumes the form
\begin{eqnarray}
    \label{DiracEq-H}
 i\sigma^1\frac{u}{\ell} \bigg ( \frac{\partial}{\partial u} + \frac{1}{2u} \bigg )&&\psi_{H} + i\sigma^2 \frac{u}{\ell}\frac{\partial\psi_{H}}{\partial x}+\\[5pt] \nonumber
+&&  \sigma^3 \left(\frac{u}{\ell}\right)^{z} \left(\frac{\partial}{\partial t} + \frac{z\ell^{z-1}}{2u^z}\sigma^2\right)\psi_H = 0,
\end{eqnarray}
where $\psi_H=A(u)^{-1}\psi$ represents the rotated Weyl spinor.


\section{Holonomy and Wilson Loop}
Since we are dealing with fermions in curved spacetimes, a natural question which arises is how the background geometry influences the interaction between the fermionic wave function and the curvature of spacetime. In particular, some of these effects can be explored through the Dirac phase method~\cite{dirac1931quantised}, which — among other applications — allows the computation of the geometric phase associated with transport phenomena in systems exhibiting non-trivial topology~\cite{berry1980exact}. To implement this approach, we assume that the fermionic spinor can be factorized as
\begin{eqnarray}\label{Diracmethod}
\nonumber \psi_{H}(t+T,u,x)&=&  \mathcal{P}\exp{ \left( - \oint_{\mathcal{C}} \Gamma_{\mu}(x)dx^{\mu} \right)}{\psi_{0}}(t,u,x),\nonumber\\
&=& \mathcal{P}\exp{ \left(- \oint_{\mathcal{C}} \Gamma_{t}(u)dt \right)}{\psi_{0}}(t,u,x),~~~~~~
\end{eqnarray}
where $T$ denotes the period of a closed path, \( \mathcal{P} \) represents the order operator and \( \psi_0(u, x)\) is the solution of the Weyl equation \eqref{DiracEq-H}, which now becomes
\begin{equation}
    \sigma^3 \frac{\partial\psi_{0}}{\partial t} + i\sigma^1 \bigg ( \frac{\partial}{\partial u} + \frac{1}{2u} \bigg )\psi_{0} + i\sigma^2  \frac{\partial\psi_{0}}{\partial x} = 0,
    \label{DiracEq-0}
\end{equation}
for $t = 0$.

According to the identity \eqref{Diracmethod}, for fixed $u$ and $x$,  when traversing a closed path $\mathcal{C}$ with period $T$, the Weyl fermion must acquire an holonomic phase $U(\mathcal{C})=\mathcal{P}\exp{ \left( - \oint_{\mathcal{C}} \Gamma_{\mu}(x)dx^{\mu} \right)}$ in analogy with the scalar Aharonov-Bohm effect~\cite{aharonov1959significance, aharonov1987phase}, where fermions acquires a geometric phase related to the presence of the electric potential. However, this effective electric potential has a non-Abelian character related to the Pauli matrix $\sigma^2$. In gauge theories, the existence of holonomy is related to the behavior of fields under closed paths, and its topological invariant can be understood as a measure of the phase acquired by the field when undergoing parallel transport. For the case addressed here, we consider the spin connections inserted in the covariant derivative used in the Weyl equation \eqref{weq} as parallel transport connections and, in this way, the captured holonomy acquires a geometric nature related to the curvature of the Lifshitz spacetime. The holonomic phase is given by
\begin{equation}
    U(\mathcal{C}) = \cosh{\left(\frac{zl^{z-1}}{2u^z}T\right)} - \sigma^2 \sinh{\left(\frac{zl^{z-1}}{2u^z}T\right)},
    \label{holonomy}
\end{equation}
or, in matrix form,
\begin{eqnarray}
    U(\mathcal{C}) = \begin{pmatrix}
        \cosh{\left(\frac{zl^{z-1}}{2u^z}T\right)} & i\sinh{\left(\frac{zl^{z-1}}{2u^z}T\right)}\\[8pt]
        -i\sinh{\left(\frac{zl^{z-1}}{2u^z}T\right)} & \cosh{\left(\frac{zl^{z-1}}{2u^z}T\right)}
    \end{pmatrix}.
    \label{matrixholonomy}
\end{eqnarray}
Note that $U(\mathcal{C})$ depends on the dynamical exponent $z$ and the scale factor $\ell$. Its associated topological invariant, the Wilson loop  \cite{cai1991neutrino, de2003loop, bezerra1992, gomes2007loop, de2025adiabaticity}, is related to the polarization of the Weyl fermion and can be expressed as \footnote{One can verify that $W(\mathcal{C})$ is independent of the tetrad choice by computing the
holonomy in the standard diagonal frame.}
\begin{equation}
    W(\mathcal{C}) =tr\left(U\left(\mathcal{C}\right)\right) =2\cosh{\left(\frac{zl^{z-1}}{2u^z} T\right)}.
    \label{wilsonloop}
\end{equation}
 This fact suggests that electronic transport in 2D Weyl semimetals can exhibit geometric phase effects, associated with the presence of curvature caused by a distribution of disclinations in their crystal lattice, as occurs in graphene in the presence of an electric field \cite{katsnelson2012graphene, lukose2007novel}. This holonomy, in particular, introduces a squeezed state through the mixing of the valley states, resulting from the modifications in the spatial distribution of the spinors due to the curvature effects. 

\section{Fermion zero mode}

In addition to the holonomy presented in Eq. \eqref{holonomy}, we also need to solve Eq. \eqref{DiracEq-0} in order to completely determine the solution of the Weyl spinor \eqref{Diracmethod}. Here we use the {\it ansatz}
\begin{eqnarray}\label{fermion_ansatz}
 \psi_{0}(t, u, x) = e^{iEt - ikx}\left(\begin{array}{c} \psi_1(u)\\ \psi_2(u)\end{array}\right),
\end{eqnarray}
which allows us to rewrite Eq. \eqref{DiracEq-0} as a coupled system of first-order ordinary differential equations involving the components of the fermionic spinor \eqref{fermion_ansatz}, given by
\bes\ben
\left(\frac{u}{\ell}\right)^{z} E\psi_1(u) &=& -\left(\frac{u}{\ell}\right)\left(\frac{d}{du} + \frac{1}{2u} + k\right)\psi_2(u),~~~~~~\\
\left(\frac{u}{\ell}\right)^{z} E\psi_2(u) &=& ~~~\left(\frac{u}{\ell}\right)\left(\frac{d}{du} + \frac{1}{2u} - k\right)\psi_1(u).~~
\een\ees
By decoupling the system above, we are led to the problem of dealing with a pair of radial equations with central potentials, expressed as
\bes\label{fac1}\ben
\left(-\frac{d^2}{dr^2}+\frac{2l}{r}\frac{d}{dr}-\frac{l(l+1)}{r^2}+V_{-}(r)\right)\hat{\psi}_1&=&\tilde{E}^2\hat{\psi}_1,~~~~~~~~~\\
\left(-\frac{d^2}{dr^2}+\frac{2l}{r}\frac{d}{dr}-\frac{l(l+1)}{r^2}+V_{+}(r)\right)\hat{\psi}_2&=&\tilde{E}^2\hat{\psi}_2,
\een\ees
where $l=1/2z$, $\hat{\psi}=u\psi$, $\tilde{E}=E/\ell^{z-1}$, the new radial coordinate $r$ is defined through the transformation $u^{z-1}du=dr$ and central potentials are
\begin{equation}\label{qp}
V_{\pm}(r)=\frac{\tilde{\kappa}^2}{r^{2\alpha}}\pm\frac{\alpha\tilde{\kappa}}{r^{\alpha+1}},
\end{equation}
with $\alpha=1-1/z$ and $\tilde{\kappa}=kz^{-\alpha}$. Note that since $z>1$, the values of the $\alpha$-parameter in the geometry-induced potential \eqref{qp} are confined to the range $0<\alpha<1$. In particular, when approaching relativistic regimes considering the limit $z\to 1$, we have $\alpha\to 0$ and $V_{\pm}(r)\to\kappa^2=const.$, indicating that the reduction of the anisotropic scaling between the space and time coordinates implies the elimination of fermion bound states. In extremely anisotropic setups, where $\alpha\to 1$ for $z\to\infty$, we also find scenarios with no bound states, since in these cases we have $V_{\pm}(r)\to 0$. Therefore, one can observe that the presence of well-behaved fermion bound states requires the existence of anisotropy between space and time. However, if this anisotropy is too intense, such states may not survive. This may indicate that the electronic properties of 2D Weyl semimetals can change if their associated Fermi velocity is too low. In particular, for $z=2$, we have $\alpha=1/2$ and we find a scenario with a Coulomb potential complemented by a higher-order term. 

The pair of equations (\ref{fac1}a,b) can be factorized as
\bes\label{ssystem}\ben
S_+ S_{-}\hat{\psi}_1 (r)&=& \tilde{E}^2\hat{\psi}_1(r),~~\\
S_{-} S_{+}\hat{\psi}_2(r) &=& \tilde{E}^2 \hat{\psi}_2(r),~~
\een\ees
with $S_{\pm} =-\left(\pm\frac{d}{dr}\mp\frac{1/2z}{r}+\frac{\tilde{\kappa}}{r^{\alpha}}\right).$ In Schr\"odinger form, we have
\bes\label{mqsf}\ben
\left(-\frac{d^2}{dr^2}+V_{-}(r)\right)\tilde{\psi}_1&=&\tilde{E}^2\tilde{\psi}_1,\\
\left(-\frac{d^2}{dr^2}+V_{+}(r)\right)\tilde{\psi}_2&=&\tilde{E}^2\tilde{\psi}_2,
\een\ees
where $\tilde{\psi}=r^{l}\hat{\psi}$. By defining the auxiliary function $W(r)=\tilde{\kappa}/r^{\alpha}$, one can rearrange the central potentials in Eq. \eqref{qp} as $V_{\pm}(r)=W^2\mp W'$, which allows us to factorize the fermion equations in Eq. (\ref{mqsf}a,b) as a SUSY-like structure \cite{cooper1995supersymmetry} given by
\bes\ben
Q^+Q^-\tilde{\psi}_1&=&\tilde{E}^2\tilde{\psi}_1,\\[2pt]
Q^-Q^+\tilde{\psi}_2&=&\tilde{E}^2\tilde{\psi}_2,
\een\ees
with $Q^{\pm}=\pm\frac{d}{dr}+W(r)$. The zero-mode components are determined by the pair of equations $\left(Q^{-}\tilde{\psi}_1,Q^{+}\tilde{\psi_2}\right)=(0,0)$, becoming
\begin{eqnarray}\label{zero-mode components}
 \tilde{\psi}_0(r) = \left(\begin{array}{c} c_1 e^{\frac{\tilde{\kappa}}{1-\alpha}r^{1-\alpha}}\\ c_2 e^{-\frac{\tilde{\kappa}}{1-\alpha}r^{1-\alpha}}\end{array}\right),
\end{eqnarray}
which is normalizable since $\alpha<1$, with $c_1=0$ if $\tilde{\kappa}>0$ or $c_2=0$ if $\tilde{\kappa}<0$. The existence of well-behaved zero modes indicates that the defects associated with the curvature of Lifshitz spacetime allow the emergence of zero-energy fermionic states in their vicinity. 

\section{Ending comments}\label{end}

In this work, we investigated properties of Weyl fermions in (2+1)-dimensional Lifshitz spacetimes. In particular, we found the geometric phase associated to the coupling of the fermion field with the background geometry, with analytical holonomy and explicit dependence on the curvature and scaling anisotropy parameters. We also studied the solution of the associated Weyl equation and captured an analytical expression for the fermion zero mode, derived from an underlying supersymmetric quantum mechanics. The existence of a continuous phase dependent on the curvature suggests that the dispersion relation of two-dimensional Weyl semimetals presents a direct analogy with the effect of an electric field on graphene. The holonomy obtained in~\eqref{matrixholonomy} is responsible for the mixing of the valley states and the shift of the Dirac cone. Furthermore, the analytical zero modes protected by topology indicate the possibility of edge states in the quantum Hall effect when a magnetic field is introduced, once the holonomy plays a role of a perpendicular electric field to the plane.

This is an initial study concerning fermions on geometries with anisotropic scaling invariance, and we hope that further investigations will be useful for future works on effective gravitational models and their applications in planar condensed matter systems, since there are several connections between the physics of gravitational systems and the physics of materials such as graphene, topological insulators, and 2D Weyl semimetals. Within this context, several other phenomena can be explored, such as Landau levels, Hall effect and the Aharonov-Bohm-like effect, for example. We hope that these studies will open up a range of new research opportunities and be useful for a better understanding of the physics of two-dimensional materials.

\begin{acknowledgments}
 We would like to thank Prof. Dr. Knut Bakke and Prof. Dr. Alexandre Carvalho for the rich discussion which helped us in this paper. G. Q. Garcia would like to thank Fapesq-PB for financial support (Grant BLD-ADT-A2377/2024). The work by C. Furtado is supported by the CNPq (project PQ Grant 1A No. 311781/2021-7). D.C. Moreira would like to thank UEPB and the Brazilian agency CNPq for the partial financial support (Process No. 402830/2023-7). 
\end{acknowledgments}

\bibliography{biblio} 
\end{document}